# Serendipity with Generative AI: Repurposing knowledge components during polycrisis with a Viable Systems Model approach

Gordon Fletcher*, University of Salford, UK (G.Fletcher@salford.ac.uk)

Saomai Vu Khan, University of Salford, UK (S.VuKhan@salford.ac.uk)

## Abstract

Organisations face polycrisis uncertainty yet overlook embedded knowledge. We show how generative AI can operate as a serendipity engine and knowledge transducer to discover, classify and mobilise reusable components (models, frameworks, patterns) from existing documents. Using 206 papers, our pipeline extracted 711 components (≈3.4 per paper) and organised them into a repository aligned to Beer's Viable System Model (VSM). We contribute i) conceptually, a theory of planned serendipity in which GenAI lowers transduction costs between VSM subsystems, ii) empirically, a component repository and temporal/subject patterns, iii) managerially, a vignette and process blueprint for organisational adoption and iv) socially, pathways linking repurposing to environmental and social benefits. We propose testable links between repository creation, discovery-to-deployment time, and reuse rates, and discuss implications for shifting innovation portfolios from breakthrough bias toward systematic repurposing.

## 1. Introduction

For a world facing multiple simultaneous crises including a climate emergency, armed conflicts, so-called "culture" wars, pandemics and rapid technological shifts, organisations must innovate to keep pace with the variety of novel problems that are continuously emerging in this complex environment. The ability to reuse and adapt transferable knowledge components such as models, frameworks or patterns can accelerate the innovation process (by applying these reusable building blocks) and reduce redundant effort (by using already tested building blocks) during this persistent polycrisis. Knowledge sharing and reuse are widely recognised as essential for innovation, both within organisations and across broader ecosystems (Abu Sa'a & Yström, 2025). Innovation literature emphasises these benefits of repurposing as a strategic approach to addressing grand societal challenges, particularly during periods of uncertainty (Omezzine et al. 2021; Liu et al., 2021). This type of reuse can leverage serendipity within an organisation and productively channel the capabilities of artificial intelligence to accelerate innovation while still attending to the legitimate concerns of economic, social and environmental sustainability. Rather than waiting for breakthroughs with uncertain arrival times, organisations can address near-term needs by repurposing existing knowledge and artefacts. We treat repurposing as a designable process involving the identification, extraction and recombination of components rather than an accident of discovery. This work contributes further to this existing body of literature by focusing on the beneficial role of GenAI in facilitating knowledge repurposing in ways that can be compatible with the existing systems in an organisation for getting things done. The core feature of generative AI and its large language models is the ability to recognise patterns and extrapolate. This capability is of significant value in analysing texts and recognising the sometimes subtle and complex regularities that are found in the vast variety of human expression. We argue that exploiting this capacity for pattern recognition and synthesis can draw out useful insights from specific texts as reusable and repeatable knowledge components. This application is a more productive use than endeavouring to replicate human creativity through expressive works such as art, poetry or music (Anantrasirichai and Bull, 2022). While creative outputs such as GenAI art, poetry or music have attracted extensive popular critique (d'Amato et al., 2025; Goetze, 2024; Xie, 2025), the scholarly attention towards GenAI's productive capabilities in terms of knowledge management and innovation contexts remains comparatively less explored. As a contribution to this debate, we present an approach to knowledge repurposing that takes GenAI's capacity to extract statistical patterns from large corpora as "a crystallized form of collective human knowledge" (Lindley et al., 2024) and applied it to finding the reusable elements of knowledge that can be found in individual academic papers. In other words, this work exploits current publicly available pattern detection technology to find knowledge patterns.

Earlier research (Kim et al., 2004; Haefliger et al., 2008; Narasimhan and Reichenbach, 2015; Ohta et al., 2015) recognised that knowledge workers, and particularly software developers, already often engage in significant content reuse (through a "copy-paste" approach to the assembly of text and code). Rather than be critically dismissive of these practices, we see GenAI as a way to improve the quality and accuracy of the approach. By advocating for the conscious improvement in this tacit, largely unspoken but commonly used aspect of knowledge work (Huang et al., 2022) organisations can focus on the practical "missing link". It is critical to find the sorely needed productivity gains that effectively breaks out of the existing "steady-state" of activities to ensure their continued existence in polycrisis economies. In this way, the polycrisis is not just a threat but can act as a catalyst that harnesses overall uncertainty as an opportunity to innovate and bring about long-term improvement (Klyver and McMullen, 2025). We are seeking ways to enable organisational systems to break out of their current situation - their homeostasis - and address the variety (of challenges) presented by polycrisis without threatening their overall viability (Beer, 1985, p. 17), by minimising the costs of introducing new ideas and transducing concepts across the organisation.

Generative AI offers a means to balance these tensions. As Krakowski (2025, p.10) observes, “In the solution-search stage, GenAI can facilitate the exploration of cognitively distant opportunities by generating new combinations of ideas and perspectives” which can help individuals and organisations to “move beyond existing or conventional mental models and domain constraints”. Building on this, our argument is that Generative AI can act as an “automated broker” of knowledge within an organisation, identifying reusable components found in previous works and structuring these components to be used across the organisation in new, even innovative, ways. We see this capability as being even more valuable in times of high uncertainty or crisis, when resources and time will be limited and known or proven insights are preferred for application into new domains as they represent less risk. The core technical advance offered by GenAI for this type of work is high quality, accurate, needle-in-a-haystack pattern-matching and this is precisely the capability being exploited. Organisations can apply this capability to introduce a heightened level of knowledge agility by repurposing existing and documented intellectual capital for application to emerging problems. We adopt the Viable Systems Model (VSM) (Beer, 1979, 1981) as the core theoretical lens to better understand and explain how AI-supported knowledge reuse contributes to innovation and systemic resilience. As a model for organisational cybernetics, the VSM describes the necessary recursive subsystems needed to maintain an organisation’s viability against complex and changing environments (Beer, 1979; Beer, 1981) . At its core, the VSM provides a framework to interpret how an organisation can manage internal knowledge (as a resource) and adapt it to the variety of unexpected external changes that occur during polycrisis. This original purpose mirrors the dynamic at play in repurposing knowledge for innovation. Using VSM, we argue that GenAI can enhance an organisation’s ability to respond to external change and uncertainty (the core function of what VSM describes as “System 4”) by drawing on insights distilled from existing knowledge assets produced internally or by other knowledge-generating institutions such as universities, i.e. the outputs of the “System 1” of operations in the VSM.

Empirically, we demonstrate these ideas through the experimental case of an academic department as a knowledge-producing system and itself part of the wider visible system of a university. We used a novel generative AI tool to analyse 206 academic papers produced over five years in one department, extracting candidate reusable knowledge components (e.g. theories, methods, frameworks, patterns, best practices). We then categorised and quantified these components to understand what types of reusable knowledge are most commonly identified. The findings provide a concrete illustration of an extensive inventory of knowledge components that an organisation unknowingly accumulates over relatively brief periods of time, and how AI can unlock this specific inventory for further use and exploitation. While a university department is expected to actively and consciously engage in this type of knowledge production work, it is a meaningful case for wider understanding and application. The volume of reusable components will be larger than in organisations not focused on producing knowledge, but as universities' work should be public and available to all, it offers other organisations the potential to also exploit this rich stream of reusable knowledge components in an accessible way. As a test case, universities also provide a setting where the structure and separation of organisational subsystems, as they are conceptualised in the VSM, are particularly visible and well-documented. In effect, universities tend to build internal structures that closely mirror the configuration of the five linked subsystems of the VSM.

In the following sections, we explain the theoretical grounding for this work through the Viable Systems Model and how it informs our understanding of knowledge repurposing. We then detail our methodology, including the collected dataset and the process of AI-based extraction. From this data we present our findings, including descriptive statistics on the frequency of different component types and examples of the components identified. In the discussion, we interpret these findings: illustrating how AI-assisted knowledge repurposing can directly lead to serendipitous insights, bolster innovation in uncertain times and contribute to sustainable innovation practices. Finally, we offer conclusions about the role of AI in enabling innovation through knowledge reuse, and discuss the implications for both research and practice within the viable systems of organisations.

This paper makes four contributions aligned to the "Repurposing for Innovation". We theorise planned serendipity as an organisational capability by articulating how generative AI reduces transduction costs between operational knowledge generation and strategic exploration which increases the chance that useful repurposing opportunities are surfaced. We show how generative AI can extract and structure reusable "knowledge components" from existing texts which amplifies variety for the subsequent recombination and acceleration of discovery-to-deployment cycles. We explain how this approach is especially valuable during periods of high uncertainty, when the rapid recombination of available artefacts is preferred over waiting for breakthroughs. Finally, we show how these components can encourage reuse and upcycling of knowledge to support a reduction in the environmental and social costs of innovation. Taken collectively, these contributions offer an operational blueprint that is grounded in the Viable Systems Model to build organisational pipelines for repurposing already available knowledge into responsive and relevant innovations.

## 2. Theoretical framing: the Viable Systems Model and knowledge repurposing

The Viable Systems Model is a framework for understanding the structure and governance of any organisation. Specifically it describes a viable organisation, one that can survive and thrive in uncertain and changing external environments.

According to VSM, a viable system is composed of five interacting subsystems (mundanely labelled as Systems 1–5) that are recursively organised (Figure 1).

In summary, System 1 units are operational elements, Systems 2 and 3 handle internal coordination and control, System 4 scans the environment and plans adaptation and System 5 defines identity and policy. For our discussion it is possible to recognise how each of these subsystems make different contributions and receive different benefits from the action of knowledge repurposing (Table 1).

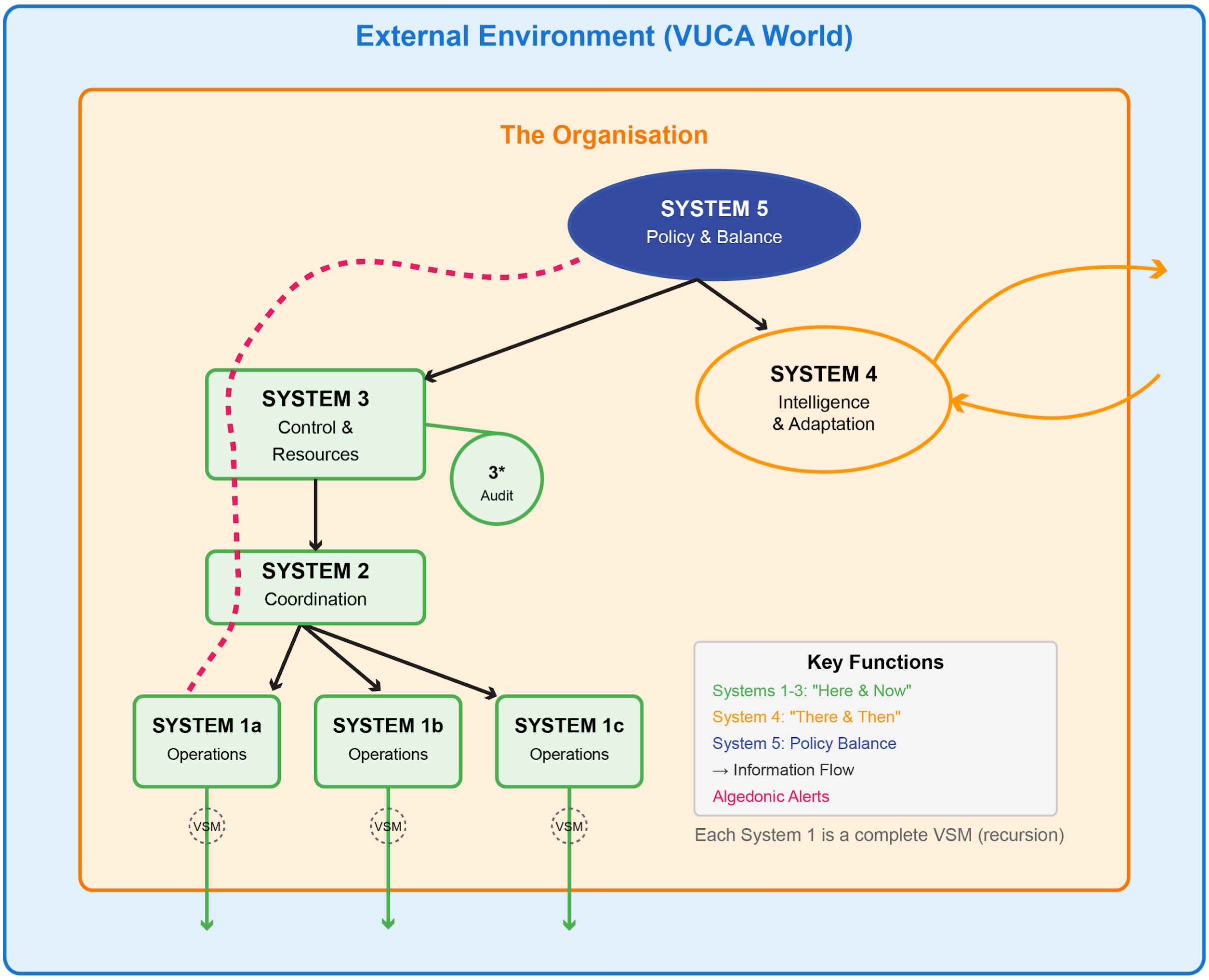

Figure 1: A simplified diagram of the Viable Systems Model

| **VSM Subsystem** | **Relation to AI-assisted knowledge repurposing** |
|---|---|
| System 1 (Operations) | Extraction of reusable knowledge components from operational outputs. |
| System 2 (Coordination) | Highlights duplication of effort (and components). Aids the coordination for maximum reuse. |
| System 3 (Control/Resources) | Provisions the AI system. Supports structured documentation and controlled dissemination of components. |
| System 4 (Adaptation) | Rapidly identifies relevant internal knowledge for adaptation to external change. |
| System 5 (Policy/Identity) | Reinforces an organisational culture that encourages reuse and learning. Promotes the value of being a learning organisation. |

Table 1: The Viable Systems Model subsystems and their relationship to GenAI

Building on Ashby's Law of Requisite Variety, Beer emphasises the need to maintain homeostasis by moving between phases of amplification and attenuation of variety within the system to meet the challenge of external variety (Beer 1979, p. 83-122). It is in this need to maintain organisational homeostasis that the rationale for innovation is also located. With large-scale change in the external environment comes the need for increased capability to undertake internal change. Without an increased capacity for innovation being introduced into the system, the ability to maintain the homeostasis of the system would become increasingly challenged. System 5 maintains this balance by providing overarching direction, ensuring decisions coming from Systems 3 and 4 remain aligned with the organisation's core mission, vision, identity and values (Beer, 1979, p.259-263). The top-level view of strategic management emphasises a further requirement to the overall organisational need for innovation. Strategic management cannot become drawn into the details of specific activities or operations, and this means that innovation must be understood within each of these subsystems in different ways. The varying needs for understanding within each subsystem requires the process of transduction. That is the translation of knowledge and meaning across organisational subsystems. A well-designed transducer preserves the complexity (or "variety") of the original signal, allowing the receiving subsystem to act effectively with the information (Beer, 1981, p.367). While transduction itself is essential for system viability, it incurs costs, both cognitive and operational, as existing knowledge must be reformulated for different contexts. These costs, often invisible and undervalued, can lead to information becoming distorted, delayed, or completely overlooked. "Imperfect" transduction represents a significant barrier to knowledge reuse, and by extension, to innovation. If these costs are not explicitly addressed through effective methods of abstraction and onward communication throughout the organisation's subsystems then the ability to maintain homeostasis during times of heightened external variability becomes difficult.

We interpret knowledge repurposing within an organisation using the VSM framework. Through this perspective, the various knowledge outputs of an organisation (which could include research papers, reports or processes) are products of System 1 units. Typically, each operational unit generates knowledge to solve very specific problems at hand, often without any intention or need to share or reuse that knowledge elsewhere at a future point in time. Over time isolated pools of knowledge components will accumulate internally including ways of working, frameworks for actions, tools to do the job, preferred or best practices and other similar artefacts of the organisation. However, these valuable components generally remain siloed and largely tacit, limiting the organisation's ability to leverage any of this knowledge into new contexts. As Beer (1979, p. 227) explains, System 4 must engage not only with the external environment of the organisation as a whole, but also with the environments and outputs of each System 1 unit. Thus, for an organisation to be innovative and resilient to external change, its System 4 should not only draw in “packaged” new knowledge from external sources, but it should also capitalise on the internal reservoir of knowledge when addressing new challenges. In practice, this means an organisation should be able to identify, abstract, and redeploy useful knowledge elements from experience (System 1 outputs) to meet emerging needs that are identified by System 4. The pathway between System 1 and System 4 is challenging and echoes the challenge of Solomon’s paradox. In other words, it is often difficult for organisations to recognise their embedded internal wisdom, and they will often favour easily accessible external expertise instead.

Knowledge repurposing is part of the solution to this Solomon’s paradox and a means to successfully achieve transduction between Systems 1 and 4 (and 5) within the VSM. Mapping knowledge repurposing actions to each of the VSM subsystems describes an idealised situation for a more tightly integrated organisation.

- System 1 (operations): Produces knowledge components that are contextual linked to specific activities and projects.
- System 2 (coordination): Reducing duplication of effort by coordinating the reuse of knowledge components across the organisation. This is where many organisations lack a strong mechanism. A consequence of low levels of coordination is a constant process of “reinventing the wheel”.
- System 3 (control and resources): Can facilitate knowledge sharing infrastructure by ensuring that knowledge components are first captured, then documented and finally made sharable in ways that are meaningful outside the original specific System 1 context and suitable for each subsystem.
- System 4 (intelligence and adaptation): Scans the external environment and identifies the need for change innovation. System 4 should also be able to query internal knowledge to find suitable ideas or tools that can be reused to address external variability. This capability requires a linkage between System 4 and System 1 that is not overburdened by the overheads of successful transduction. The need for this internal relationship places additional pressures on Systems 2 and 3.
- System 5 (policy and identity): Sets the culture and expectation to encourage learning from past projects and reusing proven solutions. Efforts to promote knowledge reuse as a normalised expectation encourages action in the other systems to undertake each of their own specific tasks. There is a cautionary need to minimise the amount of information flowing to System 5 regarding the specific reuse of knowledge to avoid adding unnecessary transduction burdens across the system.

Through this idealised model, GenAI can be seen as an enabling transducer that strengthens the capability to link System 1 and System 4 in ways that can be productive while minimising the multiple translation costs that the process normally entails. By rapidly processing large volumes of internal documents (System 1 outputs) and extracting patterns or other reusable knowledge elements, the GenAI then assists System 4 with a potentially rich menu of innovations expressed in a way that can be immediately reused. The variety of internal organisational knowledge, which is often so overwhelming that human-monitoring and sense-making is impossible, becomes tractable when handled in this way. This augmentation of existing organisational intelligence enhances the overall system's ability to remember, recognise and recombine its prior knowledge in novel ways, improving its viability and resilience to external variety. A response of this type directly addresses a key challenge in knowledge management: effectively integrating internal and external knowledge for adaptation (Grant, 1996; Cristache et al., 2025). This observation is also consistent with a core tenet of the Viable Systems Model, and expressed as Ashby's Law of Requisite Variety, in which an organisation utilising AI technology in this way can better match the complexity of external problems by calling upon a wider variety of internal solutions.

We argue that organisations leveraging GenAI for knowledge repurposing will exhibit greater innovation agility and resilience. Prior studies have hinted at the power of AI across its many different forms in knowledge work. Lindley et al. (2023) observes that generative AI builds on the "well-established practice of content repurposing" in knowledge work and can support it in new ways. For innovation management, generative AI tools can enable rapid large-scale "technological recombination" (Krakowski, 2025) by exploring potential multiple combinations of ideas programmatically, thus "accelerating the pace of discovery" (Calvino et al. 2025). These perspectives also echo the VSM position that having and testing diverse internally generated options is the key to successful adaptation. The onus and greatest benefit for this work is found in System 4 of an organisation, or more specifically, the function that assumes the responsibilities of System 4. As a result of these closely interwoven threads that we have set out, this work sits at the intersection of organisational cybernetics and innovation management by hypothesising that the AI-assisted extraction of reusable knowledge components will significantly enhance an organisation's capacity to undertake recombination-based innovation. The remainder of the paper examines this proposition empirically and discusses its implications within the current world of polycrisis.

In summary, we view the VSM as a key mechanism for conceptualising and positioning the role of repurposing within an organisation. In our setting, System 1 generates distributed operational knowledge, System 3 governs and allocates resources, System 4 scans and recombines and System 5 sets policy/identity. We define transduction costs as the cognitive and coordination effort required to translate any System 1 artefacts into forms usable by Systems 3, 4 or 5. Generative AI acts as a variety amplifier (increasing the accessible set of components) and a cost attenuator (lowering the effort to move artefacts across the systems), in effect, facilitating the transduction process by drawing out knowledge components from System 1 and making them intelligible and usable within other systems. Using this perspective we derive four propositions:

- P1 - Variety amplification: Lower transduction cost via AI-assisted component extraction increases the accessible variety in System 4 raising the rate at which repurposable opportunities can be identified.
- P2 - Planned serendipity: Organisations that routinise processes of cross-domain component search and curation convert serendipitous discoveries into planned capability, measured by higher reuse-rate and shorter time-to-solution.
- P3 - Uncertainty advantage: During periods of high uncertainty (polycrisis conditions) AI-assisted recombinations can produce greater innovation throughput than approaches reliant on solely de novo invention.
- P4 - Sustainability outcomes: Mapping specific components to organisational environmental and social objectives improves sustainability performance through the reuse and upcycling of organisational knowledge.

We revisit these propositions in the later discussion to synthesise empirical patterns and boundary conditions.

# 3. Methodology

## 3.1 Research context and data (and reuse)

We conducted an empirical study in the context of an academic department at a large university, examining its formal knowledge outputs over a five-year period. The department (anonymised for this study) produces research in various fields of business and management. This setting is appropriate because academic papers are an assumed rich source of explicit knowledge that often contain generalisable knowledge components such as frameworks, models, theories, and methods that could be repurposed. With the encouragement for open publishing and open research (Barczak et al., 2021), the outputs of all academic departments can be expected to be rich sources of knowledge components using the same approach that is described here. Consequently, the method presented also acts as an indicative description for non-academic organisations wanting to take advantage of this rich seam of available materials, mining specific areas of knowledge most relevant to their own activities. Although not explored in detail, this opportunity could be used in all forms of organisations and would be particularly valuable to resource constrained voluntary and charitable organisations as well as all levels of government.

We assembled a convenience sample of 206 academic publications authored by faculty in the department between 2021 and 2025. We focused on journal articles as the primary means of research expression used across the broad range of subjects. We excluded editorial notes and other artefacts that lack sufficient methodological or conceptual detail for component extraction. This sample covers a broad range of topics and research projects conducted by faculty and was representative of the department's formal knowledge output in that period. The subject areas covered by the sample and a year-by-year breakdown of publications is shown in Figure 1.

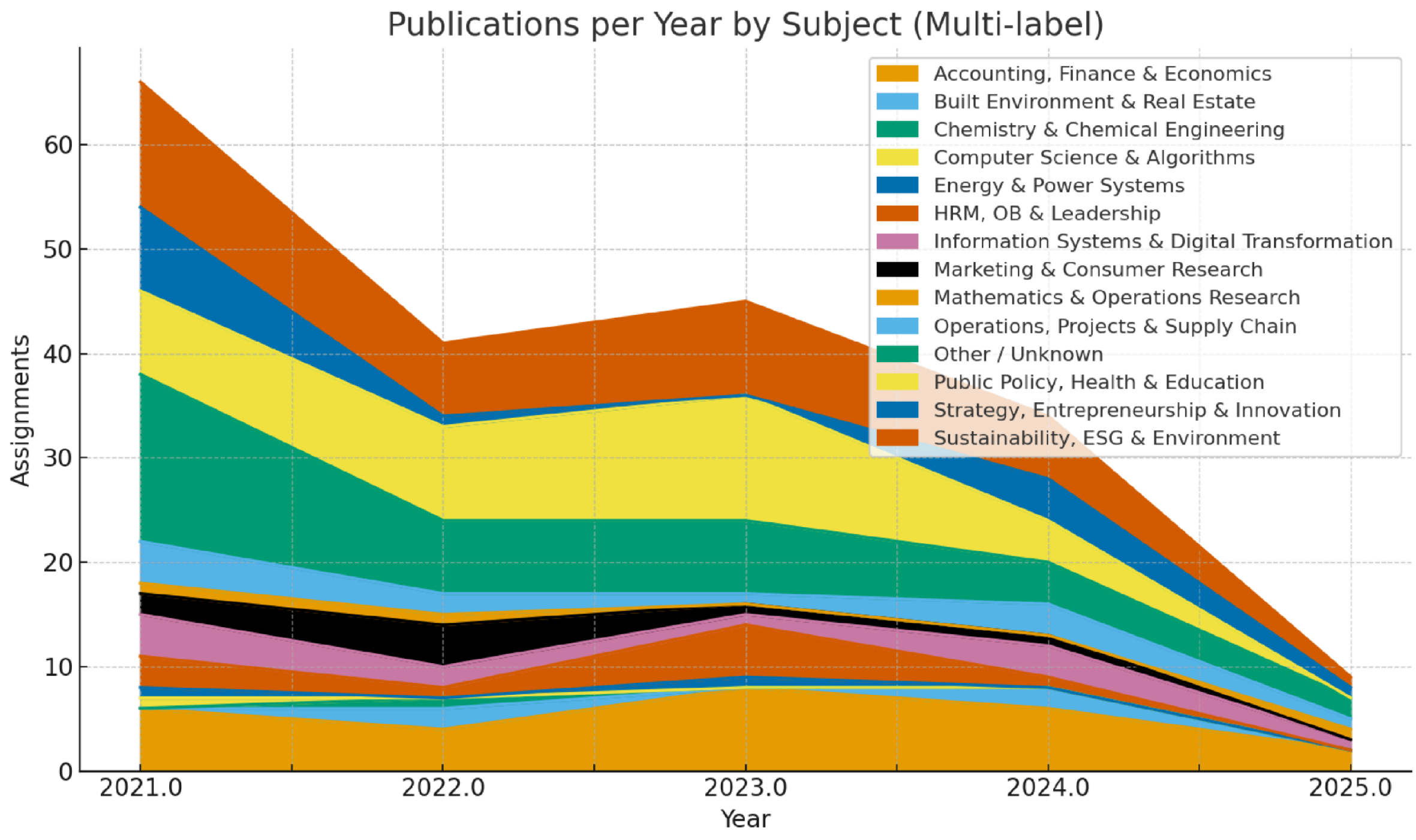

Figure 1: Paper assignments by subject by year. Totals is greater than the total sample as some papers were assigned to multiple subject areas.

Each paper in the sample was obtained in full-text (Adobe PDF) form. The metadata was recorded, but our primary interest was in the content of the papers and the identification of reusable knowledge components. "Knowledge component" included, for example, frameworks, models, patterns, checklists, best-practices or hypotheses. Essentially, we set out to extract pieces of knowledge that would have potential use beyond the immediate context of the paper for adaptation elsewhere. While the use of a convenience sample of this type produces some limitations for broader generalisability, this approach was justified pragmatically by the richness of explicit knowledge produced by the academic publications, providing an ideal initial testbed for demonstrating the efficacy of our AI-assisted extraction approach.

Our process pipeline proceeded as follows, we parsed PDFs to text and applied prompt-based extraction to identify knowledge components (definitions used are in Appendix A - system prompt). We used ChatGPT 4o with temperature 0.2, max tokens 3,000 and deterministic decoding. Output records captured component type, title/label, description, total knowledge component count for each paper and local textual evidence (span and citation). We stored the results in a structured repository for analysis and exemplification. Prompts (system and user), hyperparameters and parsing settings are provided verbatim in Appendix A.

## 3.2 Generative AI extraction tool

To identify knowledge components systematically, we developed a purpose-built generative AI tool with a web-based interface. The interface provides a convenient and consistent approach to evaluating multiple papers at the same time [DOI: 10.17866/rd.salford.30750977]. The tool leverages a large language model (LLM), specifically OpenAI's GPT4o API (or application programming interface), with a comprehensive custom prompt design to process academic texts and extract structured information about knowledge components. We iterated the prompt to ensure it produced consistent responses in style and structure. The final prompt used for all of the papers was split into two elements, the system prompt and the user prompt. The system prompt establishes the overall context for the request to the LLM, it explains to the LLM what persona it should assume and the way it should form its responses, while the user prompt sets out the specific task that is to be performed. The full text of the system prompt is provided in Appendix A.

The user prompt said,

> "Based on the provided document identify the key reusable concepts. Provide a title for the concepts (N/A if none), identify which type of concept has been discovered (N/A if none), provide a paragraph describing the specific concept discovered. The output should be formatted specifically as (with two newlines between each part of the output) with multiple concepts separated by a line e.g. ----------
>
> **Title**
>
> **concept**
>
> Paragraph description".

For the systems prompt, there was a need to define a taxonomy of knowledge component types. An initial list was drawn from knowledge management and wider relevant literature. This taxonomy contained the categories of Template (Braun et al., Cabanillas et al., 2018), Checklist (Brillinger et al. 2020; Guenther et al., 2023), Scorecard (Mio et al., 2022; Yang and Lee, 2020), Model (Chen et al., 2021; Ronaghi, 2021), Format (Holtzapple et al., 2024), Heuristic (Reijers and Limam Mansar, 2005; Frank et al., 2020), Hypothesis (Scientific or general) (Leatherbee and Katilla, 2020; Chang et al., 2024), Best Practice (Szulanksi, 1996; O'Dell and Grayson, 1998; Bergek and Norrman, 2008), Pattern (Alexander et al., 1977; van der Aalst et al., 2003), Toolkit (Von Hippel and Katz, 2002; Hermans, 2012), Theory (Scientific theory or conceptual theory)(Sutton and Straw, 1995; Weick, 1995,

Gregor, 2006; Colquitt & Zapata-Phelan, 2007), Framework (Jabareen, 2009; Gregor and Hevner, 2013; Teece, 2007), Principle (Fayol, 1949; Grant, 1996; Canbay and Akman, 2023) and Paradigm (Burrell, 1996, 1997; Burrell & Morgan, 1979; Kuhn, 1962; Shepherd & Challenger, 2012). We then did an initial draft of each term ensuring not to self-reference the concept in the definition or to duplicate the same (or similar) definitions under the different headings. We also attempted an initial hierarchy of the concepts to further ensure a separation of concepts from one another. We then utilised the Claude Sonnet 4 chatbot with the deep research function enabled to assist in refining these definitions. The chatbot was asked to check the logic of the definitions and to identify whether any other forms of knowledge component should be included. The prompt used said, "Consider the list of definitions below. Based on available insight and intelligence make suggestions for improving the definitions and identify if there are any further concepts or terms that should be added to this list." The result of this chat was used as the basis for the prompt that was included into the coded solution that utilised the ChatGPT API. A further query asked, "Based on the new definitions, produce a list of each concept and whether they fall into the category of representational tools, methodological approaches, epistemological categories or any further classification. Use this classification to broadly order the concepts from most specific to the broadest, most encompassing concept." The results of this prompt confirmed that the GenAI tool was interpreting the request correctly and appropriately defining the knowledge components (Table 2).

| **Classification by Category** |
|---|
| Representational Tools<br>● **Template** - Structured format for creating standardized outputs<br>● **Model** - Abstraction representing key aspects of systems/phenomena<br>● **Format** - Standardized structure combining multiple templates<br>● **Scorecard** - Assessment tool against defined criteria |
| Methodological Approaches<br>● **Checklist** - Systematic verification tool<br>● **Heuristic** - Rule-of-thumb for practical guidance<br>● **Toolkit** - Collection of independent tools/methods<br>● **Framework** - Structured architecture with organizing principles<br>● **Pattern language** - Coherent network of interconnected patterns |
| Epistemological Categories<br>● **Scientific hypothesis** - Testable explanation using scientific methods<br>● **Hypothesis** - Testable, tentative explanation<br>● **Best practice** - Proven technique with superior results<br>● **Pattern** - Proven solution to recurring problems<br>● **Scientific theory** - Theory developed through scientific methods<br>● **Theory** - Well-substantiated, comprehensive explanation<br>● **Principle** - Fundamental truth across multiple contexts<br>● **Paradigm** - Fundamental assumptions shaping perception |
| Meta-Conceptual<br>● **Meta-pattern** - Pattern describing how other patterns evolve/combine |
| **Ordering from Most Specific to Most Encompassing**<br>1. **Template** - Most concrete, specific structural tool<br>2. **Checklist** - Specific verification instrument |
| 3. **Scorecard** - Specific assessment instrument<br>4. **Model** - Specific representational tool<br>5. **Format** - Specific structural combination<br>6. **Heuristic** - Specific decision-making shortcut<br>7. **Scientific hypothesis** - Most specific knowledge claim<br>8. **Hypothesis** - Broader knowledge claim category<br>9. **Best practice** - Proven specific approaches<br>10. **Pattern** - Proven solutions with broader application<br>11. **Toolkit** - Collection of independent methods<br>12. **Scientific theory** - Well-established domain explanations<br>13. **Theory** - Broader explanatory frameworks<br>14. **Framework** - Structured methodological architecture<br>15. **Pattern language** - Interconnected solution systems<br>16. **Meta-pattern** - Patterns about pattern behavior<br>17. **Principle** - Fundamental truths across contexts<br>18. **Paradigm** - Most encompassing - fundamental assumptions shaping entire fields of understanding |

Table 2: Claude Sonnet 4 output ordering and categorising knowledge components

The generative AI approach was chosen over a purely manual content analysis for efficiency and consistency. Manually reading 206 papers in detail to identify and classify knowledge components would have been prohibitively time-consuming. In many respects, this is exactly the point of this paper. While knowledge is often available to an organisation it remains “locked in” by being held in the specific format of its original purpose. The likelihood of reuse is generally low because of this operational barrier. The GenAI tools processed all of the papers in a timeframe measured in hours, including the time taken to manually double check and confirm that the papers being examined were valid candidates. The GenAI pattern recognition also offered a key advantage in being able to recognise patterns that were not always obviously labelled or intended to be read as a reusable knowledge component. While a human reader might notice the papers that explicitly state, “We propose a [new framework/model] for…” and reliably extract that insight, most papers are less evident in their definition of knowledge components. Some papers even make the knowledge component contribution directly in the title, for example, “Best Practices for Leveraging BDA in Retail Logistics”. But others were much more circumspect. The paper, “Gender-Diverse Audit Committees and Non-Audit Fees: A New Perspective” is identified by GenAI as containing two patterns and a framework, but the only mention of either word is in the section title, “Theoretical Framework”. In contrast, GenAI stated that, “This study identifies a pattern in the relationship between gender-diverse audit committees and non-audit fees. It suggests that female directors on audit committees tend to reduce non-audit fees, which is associated with an improvement in audit quality. This pattern highlights the role of gender diversity in enhancing audit quality by potentially mitigating the negative effects of non-audit fees on auditor independence. The presence of female directors, known for their better communication skills and ethical behaviour, provides a unique perspective on the mixed evidence regarding non-audit fees and audit quality. This pattern can be applied to other corporate governance contexts where gender diversity might influence financial oversight and decision-making processes.” Even with this small example, it is clear that the tool can greatly accelerate and automate the extraction process and provide a structured output for quantitative analysis.

## 3.3 Analysis

Our analysis of the GenAI results are presented in two parts: i) Descriptive quantitative analysis of the extracted components, and ii) a qualitative illustration of what these components look like, to ground our discussion with specific concrete examples. For the quantitative aspects of our analysis, we compiled all the extracted components from all the papers (each entry included a paper id, citation, filename, concept count, paper title, reusable component type, and a description of the discovered component. This resulted in 711 component entries [doi: ].

We then derived descriptive statistics for the,

- Frequency of each component type overall showing how many frameworks, models and other components in total.
- Number of components per paper to determine how many reusable ideas on average a paper can or could contribute.

We present the key summary statistics in the next section, including a table of the most common component types identified. We also note underlying patterns. For the qualitative aspects of analysis, we select a series of illustrative examples of reusable knowledge components that the GenAI identified. For instance, one paper introduced “Dooyeweerd’s aspects as a framework for interdisciplinary research” which our GenAI tool correctly flagged as a framework and provided a concise description of the entire paper. Another paper offered a checklist for evaluating information system security which was identified as a checklist. We describe these examples to show the variety of components that can be discovered in a relatively modest sample size and how these might be repurposed within the organisation or (more likely) by other organisations. It should be noted that our method, while systematic, has some limitations. Any knowledge component heavily located within the figures of a paper would be missed – the current approach assumes that any visual element is a supporting aid to fuller text-based commentary. Our taxonomy may not capture all the nuances found in all papers, particularly in subject areas sitting more broadly outside business and management studies, such as psychology. Although generic, the defined knowledge components may struggle to distinguish a “model” from a “metaphor” if the author(s) describe a metaphorical model. There is also the underlying risk of hallucination from the GenAI’s Large Language Model. This was controlled by setting the consistency temperature for all the queries, the value that determines how much the model would invent components to satisfy the query, to the value of 0.2. A more cautious approach would constrain the creativity of the model further by setting this value to 0. We selected 0.2 to ensure some variability in the response while still ensuring that the paper itself was the core guiding knowledge for the response being generated. However, despite these various cautions, for the purpose of identifying clear, self-contained knowledge components our approach is effective. Any minor inaccuracies in classification do not significantly affect the high-level trends or overall organisation purpose that is under discussion.

# 4. Findings: What knowledge components were found?

Our AI-assisted analysis revealed a significant volume of reusable knowledge components in the selected research output from the department. In total, the GenAI extracted 711 knowledge components from the 206 papers (an average of about 3.5 components per paper). This confirms that even a single academic department – acknowledged as a specific academic form of System 1 - can generate hundreds of potentially reusable ideas or tools in just a few years. While these values might appear to represent a broad upper limit for knowledge components that can be developed by a single System 1 in a VSM, it does also offer an indication of the baseline of already accessible components that could be utilised by any organisation (that are not being deployed). This significant wealth and volume of existing external components still presents a challenge of external variety that a System 4 in the VSM would need to be able to effectively transduce in order to internalise.

## 4.1 Types of components identified

The most common component types identified in our sample of papers were models, patterns and frameworks that together account for over half of all identified components (Table 3). This frequency suggests that researchers regularly contribute abstract representations (the models and frameworks) and recognise potential solutions for recurring problems (the patterns) in their work. All three of these components represent key candidates for future reuse.

| Component Type | Count | Percentage of Total Components | Notes |
|---|---|---|---|
| Model | 171 | 24.1% | |
| Pattern | 129 | 18.1% | |
| Framework | 118 | 16.6% | |
| Best Practice | 87 | 12.2% | |
| Checklist | 51 | 7.2% | |
| Scorecard | 30 | 4.2% | |
| Toolkit | 29 | 4.1% | |
| Hypothesis | 22 | 3.1% | |
| Heuristic | 14 | 2.0% | |
| Principle | 13 | 1.8% | |
| Other | 39 | 5.5% | Template, Algorithm, Format and other components with less than 1% components individually |
| Unlabelled | 3 | - | Concept x 2 and configurational approach |
| N/A | 7 | - | “No reusable concepts were identified in this document” |

Table 3: Frequency of Reusable Knowledge Component Types (n = 721)

Several observations can be made from Table 3. Models are the most frequently identified type of component. These include theoretical models that explain a phenomenon as well as occasionally computational or mathematical models. The high incidence (171 instances) indicates that most papers contained at least one model and sometimes multiple models with two or three variants introduced by a single paper. The large number of models is not surprising for a set of academic papers given that the prompt expressly included the statement that, “A model is an abstraction that represents key aspects of a system, phenomenon, or concept for the purpose of understanding, analysis, or prediction” as part of the prompt. This abstraction is exactly what the vast majority of academic papers set out to achieve. Models inherently are abstractions of knowledge and set out to simplify reality to highlight the most important elements or relationships that have been discovered in a phenomenon. This makes any identified model a prime candidate for reuse in other studies as well as in non-academic contexts. The fact that the prompt provided the options of a model as the fourth knowledge component definition in the original system prompt also reassures that the GenAI did not simply prioritise the discovery of the first definition it encountered in the prompt. The application of the method used in this paper against a corpus obtained through a different organisational context, for example in an organisation other than an academic department, would produce a different weighting of components than those presented here. This is one clear opportunity for future research in the field of knowledge components and their reuse.

Patterns were the second-most common with 129 instances. In this context, “pattern” refers to a documented solution to a recurring problem and mirrors the intention of design patterns in software or the idea of pattern languages first presented for architecture (Alexander et al., 1977, p.10). The department had a number of papers, found especially in the fields of information systems and organisational studies, that explicitly discussed patterns or pattern languages as a direct contribution of the work. For example, one paper presented a pattern language for collaborative decision-making; another identified patterns of customer behaviour in online platforms. Patterns are also valuable for reuse because they present what should be proven solutions that are adaptable to different contexts. By definition, a pattern has been generalised out of repeated application. The presence of multiple patterns across an interrelated corpus also represents an opportunity for the rapid development of a pattern language in a specific field or subject. Just as Alexander et al. noted (1977, pp. 14, 44), the patterns in a language are not isolated entities, but are supported by and embedded within other patterns, allowing them to be combined and compressed in various ways to produce viable solutions to wider challenges. At the time of authoring, Alexander et al. (1977) also highlighted the complexity and time-commitment required to undertake pattern identification and then fully articulate a pattern language. GenAI used in the manner described in this paper represents a key method to resolve this resource commitment bottleneck faced by earlier authors.

Frameworks (with 118 instances) were also very prevalent. Frameworks typically provide an overarching structure or architecture for addressing a class of problems. They often come in the form of conceptual frameworks with a set of concepts and a description of how they interrelate or procedural frameworks with a step-through approach to successfully doing something. In our dataset, the discovered frameworks ranged from being theoretical, where a framework integrated multiple theories to explain a phenomenon, to practical with the framework implementing a specific practice. Frameworks are often the scaffolding for further new work. Future researchers or practitioners then apply, test, and modify the described framework, rather than starting from scratch. A framework developed internally within a System 1 would be an attractive opportunity for reuse. Having been developed within the context of the organisation and its organisational culture, structure and processes, it offers a lower barrier for adoption in other System 1s. The challenge is that the framework must be transduced through other subsystems of the viable system in order for its value to be recognised and applied where needed.

Best practice components were identified 87 times. These are specific techniques or approaches regarded as having produced better results than existing or common practice and are recommended as a direct substitute. In academic literature, best practice is often described in the review of existing literature as a synthesis of the combined results or discussed as an outcome of the research in an implications section. This observation itself is an important note, in that the source of knowledge components and the capacity to reuse knowledge is not always the result of direct or primary data collection. Utilising secondary data collection - presumably conducted by a System 4 within a viable system - can itself be a form of transduction with the act of identifying best practice through synthesising existing works and translating the concept into the organisation’s own ways of thinking and working. Best practices are generally tied to specific contexts or time periods, documenting their potential and form enables mediated knowledge transfer to similar settings and provides a qualified form of component reuse.

We also found several checklists (51 components) and scorecards (30 components). Checklists are step-by-step checks or criteria lists for ensuring completeness in a process. For instance, one paper provided a checklist for evaluating ethical risks in AI deployments. This is a clearly reusable artefact of contemporary relevance for any organisations that deploy AI and want to consider the ethical dimensions of their decisions. Scorecards are a type of evaluative framework with defined criteria that include the most often-cited balanced scorecard approach (Kaplan and Norton, 1992). An example in our data was a “sustainability scorecard” for assessing project impacts. These tools are tangible and directly lend themselves to reuse in practice.

In most cases, the papers presenting scorecards make this intention explicitly clear within their titles or in the intended outcomes of the work.

Toolkits (29 components) are a smaller category of components found in the dataset. These are a collection of tools or methods that are intended for subsequent flexible usage. One paper, for example, presented a toolkit of analytical methods for market forecasting. A key distinction from components such as a checklist but sharing a similarity with pattern languages is the recognition that not all the tools can be used successfully simultaneously. Unsurprisingly, the physical toolkit provides the clearest analogy. Simultaneously hammering in and removing a nail is impossible. Using a drill on a nail is useless and hammering in a screw is poor use of the tools available. Toolkits require knowledge of when to use which tool and when specific combinations may produce suboptimal or contradictory results. Heuristics (14 components in our data), which are simple rules of thumb or decision shortcuts, appeared less frequently. The relative scarcity of heuristics directly reflects the corpus from which the components were drawn. Academic journals rarely publish this type of knowledge component and reviewers often push back to encourage representation or removal of a heuristic. A relatively small number of peer-reviewed journals would progress a work based on presenting solely heuristics. But a different corpus would likely produce a different balance of knowledge components and heuristics might be a more expected component in, for example, an organisation's process document. Nonetheless, in our data one example included a heuristic rule for when to escalate an issue in project management, a discipline where quick decision-making know-how can certainly be critical.

Principles (13 components) can be seen as labels for more fundamental underlying truths or worldview concepts. These were less common in the dataset, but one paper explicitly discussed guiding principles for digital innovation and was included as a principle knowledge component. Overall the "other" category (39 components from a range of types) included templates as structured documents or formats to be reused, specific algorithms, protocols and metrics. For instance, an algorithm for optimising scheduling was presented in one operations research paper and a "security protocol" was detailed in a cybersecurity focused paper. At this low level of frequency, it becomes apparent that individual authorial preference in description and presentation format is a partial factor in determining what components are discovered. A telling example can be found in the five algorithms identified across four papers with the common feature being that they share the same departmental author. A single metric and one protocol were found across the entire corpus.

We undertook further analysis of the results incorporating consideration for year and subject area against the type of knowledge component (Table 4). Overall this analysis shows a sparse grid with some specific preferences for component and subject such as the relationship of models to both the Sustainability, ESG &

Environment and Public Policy, Health & Education areas. Overall the annual trend and the volume of each knowledge component remains consistent.

| Year | Subject | Most used count | Most used |
|---|---|---|---|
| 2021 | Accounting, Finance & Economics | 8 | Pattern |
| 2021 | Computer Science & Algorithms | 1 | Best practice |
| 2021 | Electrical & Electronic Engineering | 4 | Model |
| 2021 | Energy & Power Systems | 2 | Unspecified |
| 2021 | HRM, OB & Leadership | 4 | Model |
| 2021 | Information Systems & Digital Transformation | 3 | Framework |
| 2021 | Marketing & Consumer Research | 2 | Framework |
| 2021 | Operations, Projects & Supply Chain | 2 | Best practice |
| 2021 | Other / Unknown | 12 | Pattern |
| 2021 | Public Policy, Health & Education | 6 | Model |
| 2021 | Strategy, Entrepreneurship & Innovation | 8 | Pattern |
| 2021 | Sustainability, ESG & Environment | 17 | Model |
| 2022 | Accounting, Finance & Economics | 3 | Checklist |
| 2022 | Built Environment & Real Estate | 1 | Best practice |
| 2022 | Chemistry & Chemical Engineering | 1 | Best practice |
| 2022 | Electrical & Electronic Engineering | 1 | Checklist |
| 2022 | HRM, OB & Leadership | 2 | Pattern |
| 2022 | Information Systems & Digital Transformation | 1 | Best practice |
| 2022 | Marketing & Consumer Research | 3 | Framework |
| 2022 | Mathematics & Operations Research | 1 | Framework |
| 2022 | Operations, Projects & Supply Chain | 1 | Checklist |
| 2022 | Other / Unknown | 5 | Model |
| 2022 | Public Policy, Health & Education | 9 | Pattern |
| 2022 | Strategy, Entrepreneurship & Innovation | 1 | Checklist |
| 2022 | Sustainability, ESG & Environment | 14 | Model |
| 2023 | Accounting, Finance & Economics | 5 | Pattern |
| 2023 | HRM, OB & Leadership | 4 | Pattern |
| 2023 | Information Systems & Digital Transformation | 1 | Best practice |
| 2023 | Marketing & Consumer Research | 1 | Hypothesis |
| 2023 | Other / Unknown | 7 | Model |
| 2023 | Public Policy, Health & Education | 11 | Pattern |
| 2023 | Sustainability, ESG & Environment | 11 | Model |
| 2024 | Accounting, Finance & Economics | 6 | Model |
| 2024 | Built Environment & Real Estate | 1 | Format |
| 2024 | HRM, OB & Leadership | 1 | Best practice |
| 2024 | Information Systems & Digital Transformation | 2 | Framework |
| 2024 | Marketing & Consumer Research | 1 | Framework |
| 2024 | Operations, Projects & Supply Chain | 2 | Checklist |
| 2024 | Other / Unknown | 3 | Framework |
| 2024 | Public Policy, Health & Education | 3 | Model |
| 2024 | Strategy, Entrepreneurship & Innovation | 2 | Best Practice |
| 2024 | Sustainability, ESG & Environment | 4 | Framework |
| 2025 | Accounting, Finance & Economics | 1 | Best practice |
| 2025 | Operations, Projects & Supply Chain | 1 | Best practice |
| 2025 | Other / Unknown | 4 | Model |
| 2025 | Strategy, Entrepreneurship & Innovation | 1 | Best Practice |
| 2025 | Sustainability, ESG & Environment | 1 | Framework |

Table 4: Analysis of knowledge components by subject area, year and type.

Overall, the distribution in Tables 3 and 4 suggests that the department's research focused heavily on producing reusable concept artefacts (models and frameworks) and documented solutions (patterns and best practices). This indicates the potential to define a fingerprint identification of different academic departments - and System 1s more broadly. A comparative methodology between different departments offers a future research opportunity that breaks away from citation metrics to consider the types and relative frequencies of components produced by each System 1. What is being offered by the department examined are the types of knowledge components that, if shared and repurposed, could fuel innovation beyond their original presentation. It's worth noting that the presence of multiple component types in a single paper was common – for example, a paper might propose a new model and then offer several underlying principles, and conclude by offering a checklist for practitioners. One paper produced 8 components with five models, a framework, pattern and best practice and another had 7 components with two theories, two frameworks, a toolkit, best practice and pattern. A further two papers produced six components. No discernible pattern of component types emerged across these particularly rich papers. The GenAI tool was not restricted in how many components it could capture, so all the possible components are present in the data. We observe that the papers from specific authors and those focusing on harder science (such as transport optimisation or energy grid efficiency) tended to yield more components, while papers on more conceptual topics yielded fewer. To reduce the effect of authorial preference and understand the extent to which different subject areas produce more (or less) reusable components in any single work, a much larger dataset spanning multiple disciplines and organisations is required.

## 4.2 Examples of components

To illustrate what the specific components looked like in context and as output of the GenAI tool, we present a selection of examples drawn from the overall corpus:

Framework example: "Interdisciplinary Research Aspect Framework." One paper introduced "Dooyeweerd's aspects as a framework for interdisciplinary research." This framework is entirely based on Dooyeweerd's philosophy and structured research problems into a set of aspects including ethical, social, technical and other categories to ensure comprehensive analysis. The GenAI tool captured this complexity and summarised it as "A structured architecture defining categories (aspects) to analyze complex problems, ensuring all dimensions are considered (a framework for interdisciplinary collaboration)." This framework is intended to be highly reusable as stated by the single author. Any researcher tackling a complex socio-technical problem should be able to apply it to ensure all aspects have been covered in their analysis. Given its heavily philosophical roots the work sets out to codify a means for ensuring holistic thinking in research.

Model example: "Technology Adoption Spiral Model." This paper proposed a model describing how organisations adopt new technologies in a spiral or iterative pattern. It incorporates phases of pilot testing, evaluation, scaling and reassessment before looping back for continuous adaptation. The model is visually depicted in the paper but (fortunately for the GenAI tool) is also described extensively in the body of the work. The GenAI tool determined that this was a model and noted it as "an iterative model of technology adoption with feedback loops, emphasizing continuous learning." This component could be repurposed for the study of technology diffusion and could be applied by practitioners as a roadmap for introducing innovations into their own organisation. Its iterative design makes this model applicable to many organisational settings that require continuous technological adaptation.

Pattern example: "Customer Co-creation Patterns." In this paper from the marketing subject area, the authors identified several recurring patterns for successful customer co-creation initiatives including a "lead user involvement" pattern and a "crowdsourcing contest" pattern. Each pattern describes a problem and a solution along with the necessary conditions for success. The GenAI tool extracted these as multiple patterns. These patterns are clearly described in the paper and could be used directly in other organisations to implement co-creation. The identified patterns also offer the foundations for the development of a more expansive pattern language for open innovation initiatives.

Checklist example: “AI Ethics Assessment Checklist.” This paper offers a checklist of 10 yes/no questions to assess ethical readiness of an AI project and includes consideration for data bias, transparency and privacy. The checklist is clearly presented in the paper. The primary task of the GenAI was to summarise the longer description. This work is clearly intended to offer the checklist as a highly reusable knowledge component. Any AI project team could directly adopt this checklist to ensure they have assessed each of these key ethical issues. The paper is also an example of how academic research can directly present practical tools to an intended industry/sectorial audience. Although the checklist was not hidden, capturing it as a summarised reusable component allows for easier transfer within an organisation and for sharing elsewhere. This is a clear example of how the GenAI tool can help to overcome the challenge (and cost) of transduction. This is possible by abstracting the detail of the work away from a single System 1 and capturing its essence for the attention of System 4 and 5 within the organisation or for others.

Toolkit example: “Supply Chain Resilience Toolkit.” Drawn from an operations management paper and motivated by the disruptions caused by the pandemic, the work presents a toolkit of methods to improve supply chain resilience such as scenario planning tool, stress-test simulations and a communication protocol. The toolkit combines independent tools adapted from prior literature into a new bundle for integrated use that solves a current problem Our GenAI tool identified it as a toolkit and it is a clear case of repurposing existing ideas. Other organisations could reuse the entire toolkit while also being guided in reflecting on their current isolated use of individual tools.

These examples demonstrate the range of granularity found in each identified component. The components range from high-level paradigms to very specific instruments for direct use, and sometimes in a very specific context. But in all cases, the identified component is sufficiently self-contained that someone other than the original author could apply it. This is the essence of knowledge repurposing. By abstracting knowledge from its source context it can provide value in new contexts. With our research, Generative AI proved to be capable of identifying where authors had already done some level of abstraction. In doing this task, the tool creates a bridge between individual documents and the collective knowledge base of the organisation. This bridge – in the summation of long complex documents through identifying knowledge components – represents a means to connect the five subsystems found in the viable system of an organisation. Using the GenAI tool emphasised the real problem in knowledge reuse: transduction. In many cases the reusable component is clearly flagged in its own context of the individual paper, but presenting the component in the correct language for other subsystems is challenging. For academic work, that means lifting the paper out of the System 1 of research and presenting it so other organisational subsystems and the System 4 (the consciously externally facing system) in other organisations can clearly see that the component solves a problem that particular system is already facing.

# 5. Discussion

Our findings are consistent with our propositions 1 and 2. GenAI lowers transduction cost and increases access to internal variety while offering the capacity to routinise institutionalise serendipity by releasing the latent knowledge capital that remains under-exploited in existing work. Our proposition 3 - the uncertainty advantage - is also confirmed by the ability of GenAI to rapidly uncover and extract knowledge components for recombination at a pace that quantitatively exceeds *de novo* invention pathways. Finally, that direct mapping of component types to sustainability outcomes - in proposition 4 - enables the upcycling of knowledge to directly contribute to measurable environmental and social gains. In the following discussion, we consider how GenAI could be usefully applied for knowledge repurposing.

## 5.1 Repurposing with AI: Enhancing organisational intelligence

GenAI effectively serves as an “amplifier” for knowledge management. Conceptualising GenAI as an amplifier explicitly ties back to the VSM framework, illustrating how it significantly enhances the information-processing and adaptive capabilities of System 4, by presenting distilled internal knowledge directly relevant to emerging external challenges. GenAI also serves a secondary role as a “discoverer” for works where the presence of these knowledge components is not explicitly stated (or recognised) by the authors. In our own tests, the tool read through hundreds of texts, extracted components and systematised the results in a way that could take a human weeks to complete. Even more significantly, this task itself is unlikely to ever happen in an environment of constrained resources and continuous external change (even though this is exactly the reason for undertaking it). In VSM terms, the GenAI augments System 4 by providing rapid access to internal knowledge structures in an understandable way that can be matched to external opportunities or problems. The result of this transduction to a more compatible form for interpretation within System 4 offers the potential to create a more responsive and innovative system, as the organisation can quickly react to new challenges by drawing on a menu of documented knowledge components. Greater internal variety (of solutions) can more effectively maintain homeostasis against greater variety (of challenges). Traditional knowledge management systems rely on manual curation or employees remembering and consciously documenting and sharing insights. Approaches of this type are fraught with risk for the organisation as it relies on multiple manual interventions. In contrast, the GenAI tool can initiate a search for components across an entire corpus of organisational knowledge that has already been created for its intended purpose, generally found in the operations of System 1, and requires no further input from the original authors. This raises the possibility for a pro-active application of the same GenAI tool. Rather than (or as well as) accumulating a growing menu of available knowledge components, a new challenge can be addressed through an immediate response by scanning all internal research and pinpointing components that respond to the challenge. A further advantage is that a GenAI tool does not suffer from departmental silos or human cognitive biases. What can be described colloquially as the repeated selection of the same “usual suspects” and their work. One advantage of theorising this approach through the lens of the VSM is that its various subsystems do not align with the organisational structure or departments (Beer, 1981; Espejo & Gill, 1997, p.600), but rather defines them by their relationship and function in relation to the VSM itself.

A broader insight from our study is that the act of extracting components with GenAI in such a systematic and definitionally constrained way assists in revealing gaps or opportunities in the current knowledge base. For example, if the GenAI tool was to find few “metrics” in the outputs, this may be a sign that there is a gap in which developing standard metrics could be an innovation opportunity for the organisation. Analysis provides a means to reuse what is already there but it can also reveal gaps in an organisation’s strategy, highlighting what could be or needs to be created next.

## 5.2 Serendipity and unintended discovery

Repurposing knowledge inherently carries a serendipitous aspect. Identifying a component in one context can introduce an unexpected benefit in another. In VSM terms, one System 1 can readily offer new solutions to another. GenAI amplifies the chance of these “happy accidents” happening by rapidly exposing a much larger array of components to “innovators” - in their many varied forms. As the AI operates by identifying defined patterns it will find more components than any single researcher could possibly achieve. In effect, it can suggest reuse possibilities that would not be obvious to an individual because of personal biases and the subtlety of the pattern being expressed by the original author(s). Our GenAI tool identified multiple papers in different academic subject areas that all applied a similar underlying model of change, irrespective of whether it was organisational change, behavioural change or technological change. Recognising the presence of a meta-pattern of this type could inspire innovation that draws upon new combinations of these models. For example, adapting a form of change management best practice drawn from organisational studies work could help drive user adoption in a technological context. These types of connections are regularly missed because they become represented in two separate fields of publishing that few people are likely to be comprehensively reading across. This echoes the idea that groundbreaking solutions can be identified by reusing existing knowledge from one domain in new ways. GenAI increases the chances for serendipity by being able to present more parts of a solution in a consistent systematic way. The diverse set of components drawn from one academic department's papers (Table 3) displays the variety of ideas that can be readily recombined into future works. This material could then contribute directly back into the same System 1 that generated the original works or through application they could be introduced into other external organisations through their own System 4 as a result of the transduction work that GenAI has already completed. By making these actions more explicit, the AI sets up the opportunity for creative recombinations. Our empirical work stops at this point, however, it is quite possible to imagine a second GenAI tool that could take these transduced components and suggest new recombinations that are based on the parameters of a specifically prompted challenge. An opportunity for further research.

Serendipity is not just luck, it favours those with a suitably prepared database of knowledge components. An accessible AI-assisted repository, populated using the methods described here, creates a far greater chance that someone will find an existing idea that sparks a solution when a novel problem arises. This is an innovation sandbox: a collection of knowledge components from which institutional innovators can assemble new configurations. With the incorporation of additional AI-assistance that can make suggestions such as, "model A coupled with framework B could help to address challenge C" a further level of transduction is possible. For instance, our analysis identified a customer co-creation pattern originally intended for marketing contexts that serendipitously aligns closely with patterns of community engagement described in urban sustainability literature which could bring together new avenues for unexpected interdisciplinary innovation. An additional direction is the integration of external knowledge with internal serendipity. While our study focused on internal resources, this could be readily extended. The existing tool can effectively accept any correctly formatted document from any organisation. Regularly monitoring new research or shared documents from other organisations would allow a straightforward combination of internal and external components, potentially leading to serendipitous external collaborations between authors or technology transfer by repurposing knowledge across organisational boundaries with an open innovation mode of working. These possibilities, however, extend beyond the scope of the current study and are only a modest next logical step from the current results.

Our evidence suggests that serendipity can be routinised through the application of three design/repurposing levers. Through cross-domain component search and the scheduling of periodic activity, teams within System 4 can be challenged to source at least one knowledge component from an adjacent field. The tracking of reuse-rate measured as the percentage of projects using at least one external component assists in assessing the relative value of individual components and the rate in which change is occurring in an organisation. A second lever is the use of curated repositories activated through algedonic triggers - the 'alarms' that propagate rapidly through the VSM - when environmental or performance signals cross pre-defined thresholds and require urgent actions. In response, System 4 initiates a recombination sprint. Metrics such as the time from trigger to solution would show the value and impact of this level. Finally, lightweight recombination sprints undertaken in a constrained period (such as a single week) would assemble solutions composed of 3 to 5 components into a proof-of-concept. The hit-rate (the proportion of solutions that lead to adoption) would be tracked. Together, these practices can convert “lucky finds” into regular predictable throughput.

## 5.3 Innovation during polycrisis

In a period of polycrisis, organisations do not have the luxury of long R&D cycles. Constrained resources and brief time spans require quick actions that can apply existing knowledge to new purposes. The current period of polycrisis in which we are all operating refers to simultaneous and causally entangled crises of global scale and complexity (Lawrence et al., 2024, p. 2). It encapsulates current global challenges, including climate change, pandemics, geopolitical conflicts, and economic uncertainty (Lawrence et al., 2024).

Our findings suggest that having a menu of identified components is readily achievable as a general purpose “toolkit” for responding to polycrisis. While large (and multinational) organisations have sufficient resources to benefit from periods of uncertainty and chaos, smaller organisations are directly threatened by the same environment and in many respects are conceptually the target of larger organisations. Having the organisational capacity (through System 4) to deploy components rapidly supports the need to fight increased (external) variety with increased (internal) variety. By bringing elements from their own knowledge menu into play as a way of dealing with change, smaller organisations can simulate the actions of larger organisations. During the COVID-19 pandemic, many manufacturers repurposed production lines by amplifying variety in the form of their knowledge of manufacturing processes in order to produce PPE or ventilators (Ho et al., 2022), essentially reusing known patterns for manufacturing in a new context. Analogously, if an organisation faces sudden disruption, an AI-assisted knowledge base can quickly identify relevant frameworks or checklists from past research that can then be adapted to the situation (Box 1). In our dataset, we saw papers that were clear responses to uncertainties, including the supply chain resilience toolkit responding to pandemic-induced uncertainty. These papers themselves repurposed concepts, with existing tools repackaged for a newly identified threat. If an AI-assisted identification of components had been in place prior to this change, it may have accelerated these responses by rapidly identifying useful concepts.

### Supplier failure in a mid-sized manufacturer (5-day sprint)

**Trigger (Monday, 09:00):** A Tier-2 supplier collapses. Two critical parts within the organisation are at risk within 10 days.

**Components pulled from the repository**

- “Framework for Implementing Industry 4.0 Roles in PSM” clarifies who does what in digital procurement during disruption
- “Checklist for Role Adoption in Industry 4.0 PSM” rapid gap scan for PSM capabilities and handoffs
- “Smart Working Roles in Industry 4.0” (pattern) staffing pattern for cross-functional, time-boxed sprints
- “Stochastic Scenario Generation for Uncertainty Management” (Heuristic) a fast what-if sampling of lead-time and yield to size buffers and split orders.

**Day 1 (System 4 scan and System 3 allocate)**
Using the Framework System 4 brings together procurement, engineering, and operations. System 3 designates a sprint lead and sets a 1-week budget. The checklist flags two gaps in supplier risk sensing and dual-sourcing SOP

**Day 2 (design options)**
With the Smart Working Roles pattern, a small team prototypes three options: (i) dual-source from an EU vendor, (ii) re-spec to in-house machining, (iii) short-term broker purchase. The Stochastic Scenario heuristic generates 1,000 scenarios on demand uncertainty and lead-time variance to compare service-level impact.

**Day 3 (pilot decision)**
System 3 green-lights option (i) and (ii) effectively a split between 70/30 across EU vendor/in-house. Buffer sizes and safety times are tuned from the scenario outputs.

**Day 4 (execution)**
Engineering releases re-spec. Procurement places split orders. Operations schedule overtime for the in-house portion of the work.

**Day 5 (Review and scale solution)**
System 4 reviews KPIs and codifies the dual-sourcing SOP into the repository as a newly discovered pattern for reuse.

**KPIs (before/after)**

- Fill rate risk reduced by <X>% → target ≥ 95%.
- Expected stock-out days cut by <X>
- Time-to-mitigation: 5 business days
- New dual-Sourcing SOP pattern) available with evidence links

**VSM handoff:** System 1 data to System 4 (scan/simulate) through the use of the heuristic. System 3 funds the pilot and System 5 adds dual-sourcing to procurement policy.

Box 1: Supplier failure in a mid-sized manufacturer - a vignette drawn from discovered components of the constructed repository

There is a feedback loop here. Crises of all types force repurposing as they confront an organisational system with new external variety. Repurposing knowledge helps to deal with crises better by responding with an amplification of internal variety. Our empirical work shows how much amplification could be possible. Organisations that systematically leverage their knowledge will likely be more innovation-resilient with adaptability and flexibility possible in innovation processes that are (now always) under stress from polycrisis.

VSM offers the theoretical explanation. In turbulent environments, the role of an organisation System 4 becomes critical. It must be able to rapidly realign the organisation by drawing from internal and external insights with the urgency to consider any potential solution during a crisis. GenAI can feed this System 4 need with internal insights at the speed necessitated by this turbulence. Crisis compresses the time available for innovation and in response, GenAI extends the organisational ability to meet the challenge of time pressured decision-making. A practical implication of recognising such polycrisis is that organisations may invest in building AI-assisted knowledge repositories as a preparedness measure. This is a new perspective. While organisations have knowledge management systems, few use GenAI to dynamically transduct their existing knowledge in such a conscious or systematic manner. Our work suggests doing so could become a valuable competitive advantage, especially in (current) volatile times.

## 5.4 The sustainability of knowledge upcycling

Just as sustainability in manufacturing encourages reusing materials to reduce waste, sustainability in innovation involves reusing knowledge ideas to reduce redundant effort and accelerate progress. By recycling intellectual capital, organisations avoid discarding knowledge after a single use or presentation. This saves time and resources. An illustrative example from our dataset includes a “Sustainability Scorecard”, designed to measure the environmental impact of projects and exemplifying how explicit reuse can directly support organisational sustainability initiatives. This thinking also contributes to broader social innovation productivity. More problems can be solved with less total research investment when knowledge is shared in a structured way, whilst still retaining all of the intended meaning. This exactly expresses the transduction problem. Our analysis highlights many knowledge components that already encourage ongoing upcycling by explicit identification in the title. This emphasises the fundamental point of all scientific activity, the slow aggregation of knowledge that is founded on strong preceding works. The ‘recontextualisation’ of knowledge components is an essential step for knowledge reuse, where these components can be refined and improved to address new contexts (Markus, 2001).

We can map knowledge component types to sustainability objectives. Scorecards link to the measuring of emissions and waste. Frameworks can prioritise circular interventions and toolkits as well as patterns show how to implement low-waste processes. Table 5 shows exemplar components already found from our data and proposed outcome measures. This association directly connects knowledge upcycling to tangible improvements in environmental and social performance, extending repurposing beyond cost and speed metrics.

| Component Type | Sustainability Objective | Title | Proposed Outcome Measures |
|---|---|---|---|
| Scorecard | Measure emissions, waste and resource intensity | Comparative Evaluation of Recovered vs. Virgin Coagulants | kg waste avoided per project; % data coverage; audit pass rate |
| Scorecard | Measure emissions, waste and resource intensity | Energy Poverty Evaluation Indexes | kg waste avoided per project; % data coverage; audit pass rate |
| Scorecard | Measure emissions, waste and resource intensity | Triple Bottom Line Performance Measurement | kg waste avoided per project; % data coverage; audit pass rate |
| Framework | Prioritise circular interventions & governance alignment | Global Energy Security Collaboration Framework | Share of portfolio with circular pathways; % supplier tiers with targets; policy adoption latency (days) |
| Framework | Prioritise circular interventions & governance alignment | Optimal Nanogrid Planning Framework | Share of portfolio with circular pathways; % supplier tiers with targets; policy adoption latency (days) |
| Framework | Prioritise circular interventions & governance alignment | Integrated Energy System Management Framework | Share of portfolio with circular pathways; % supplier tiers with targets; policy adoption latency (days) |
| Toolkit | Implement low, waste processes and practices | Stochastic Modeling for Uncertainty in Energy Systems | Scrap rate delta%; water/energy per unit delta%; defects per 1k units delta% |
| Toolkit | Implement low, waste processes and practices | Green Finance Mechanisms for Renewable Energy | Scrap rate delta%; water/energy per unit delta%; defects per 1k units delta% |
| Toolkit | Implement low, waste processes and practices | Toolkit for Active Building Integration into Energy Systems | Scrap rate delta%; water/energy per unit delta%; defects per 1k units delta% |
| Pattern | Standardise low, waste routines & operational tactics | The 5D Evolution in Energy Systems | Time-to-pivot (days); % processes adopting pattern; rework hours delta% |
| Pattern | Standardise low, waste routines & operational tactics | Energy Trilemma in Building Energy Systems | Time-to-pivot (days); % processes adopting pattern; rework hours delta% |
| Pattern | Standardise low, waste routines & operational tactics | Smart Green Charging Patter | Time-to-pivot (days); % processes adopting pattern; rework hours delta% |

Table 5: Exemplars of knowledge components for sustainability from the current dataset.

Continuous improvement through reuse also leads to a form of knowledge sustainability. In other words, knowledge that endures and evolves to become more broadly applicable and have impact in a variety of contexts instead of being a one-off output. Avoiding the desire for one-off outputs is a persistent challenge in academic, organisational and governmental contexts where intermediate metrics, such as the publish or perish imperative, can become prioritised over being designed for reuse.

VSM places heavy emphasis on the capacity for communication between systems as being key to a successful (viable) organisation. If individual systems within the organisation develop their own (duplicated) knowledge there is a problem with communication and the capacity of the organisation to successfully transduct knowledge between each of the subsystems. Our GenAI approach specifically facilitates that communication process. In an academic context, this reaffirms the importance of open science as “transparent and accessible knowledge that is shared and developed through collaborative networks” (Vicente-Saez and Martinez-Fuentes, 2018, p. 428). The process mirrors the biological analogy used by the VSM where the emphasis is on the relationship of brain to body. In this case, the metaphor is found with nutrients cycling through a body where nothing is wasted and the outputs of one process become the inputs for another. Bringing new variety into the system for minimal resource cost is then an effective means to address the variety of external challenges being brought to bear by polycrisis.

## 5.5 The role of VSM in AI-assisted knowledge management

The VSM clearly frames the importance of knowledge repurposing, guiding the implementation of such AI-enabled systems. The knowledge repository would lie within System 3, the coordination system, with access to all operational knowledge, the System 1 produces outputs, in (near) real-time so that these insights can be transducted on demand to System 4, the system of strategy and innovation. In designing an organisational process, there is value in building a dedicated AI-assisted “knowledge transducer" that continuously scans and feeds relevant knowledge to ongoing projects and strategic planning from System 1 to System 3. Our findings also point to how organisations might measure their current knowledge through metrics such as those produced in our descriptive statistics; the average number of knowledge components per project or output and the diversity of component types. These would also contribute to deriving a reuse rate for these components over time. VSM encourages the presence of feedback loops across the system including algedonic signals that alert the system when under stress (Beer, 1979). These measures could produce a feedback metric where successful instances of reuse are then reported back as part of the metadata of the components held in a repository. This could involve tagging a component as having been, “applied in project X with these modifications.” Feedback of this type can support increasing trust and confidence in specific components that, in turn, will increase the reuse of these components over time. Over time, such recursive monitoring, assessment and adaptation support the creation of a true learning organisation – a core principle of viability in the VSM (Espinosa et al., 2023).

## 6. Conclusion

This paper shows how generative AI facilitates serendipitous discovery, enhances adaptive capabilities, leverages technological advances for knowledge recombination and supports sustainable organisational practices through systematic knowledge reuse. We present an approach for using generative AI as a tool to identify and repurpose knowledge for organisational innovation. Through a VSM theorised perspective and an empirical examination of 206 academic papers, we argue that AI can unlock under-utilised value in existing knowledge in ways that enable more rapid and effective innovation, especially in times of polycrisis.

The paper offers a series of key insights. The first is that many organisations possess untapped reusable knowledge components embedded in their prior activities. These components can be summarised as frameworks, models, patterns and other forms that, when systematically represented, can be recombined to address new challenges. Our case study demonstrates that even within a single academic unit there are hundreds of such components waiting to be “upcycled” in new and improved ways. Building on this insight, the pattern recognition and language understanding capabilities of generative AI make it ideally suited to perform this extraction and curation task at scale and pace. This is an example of a key productivity gain provided by AI. While this task could be done systematically by solely human hands and suggesting an efficiency gain, the likelihood of this realistically occurring in constrained times of polycrisis approaches zero. By applying the VSM as a lens on this situation, we see that undertaking this practice can increase an organisation’s adaptive (through System 4) and integrative (through System 3) functions and improve overall viability. In effect, AI-assistance helps the organisation to better know what it already has and presents that knowledge in the right context. This can be seen as a cybernetic enhancement to management, extending the classic view of organisations as self-regulating systems (Beer, 1979, 1981).

We conclude that AI-assisted repurposing of knowledge contributes directly to the potential for innovation in an organisation, can support the fostering of serendipitous connections leading to creative breakthroughs and embodies a more sustainable approach to knowledge management that encourages reuse and evolution over continuous re-invention. The approach presented shifts innovation from being seen as a linear pipeline of activity to a more holistic and circular model, where multiple outputs can feed new inputs in a multiple iterative way.

Organisations pursuing the approach described here require investment in technology and culture. On the technology side, deploying customised Nature Language Processing (through such tools as Python and SpaCy) and public LLM-based tools on internal document repositories is increasingly feasible with enterprise AI solutions already emerging for knowledge management (Lee et al., 2024). Our approach makes the case for tailoring the AI to extract structured knowledge as systematic “building block” components that provide a solution to the transduction problem of communicating to the different organisational subsystems. Culturally, organisations need to support an environment that encourages the “natural” documentation of key insights as part of the operational task at hand. This type of culture could be encouraged by rewarding reuse. Over time, building a large library of knowledge components could become a source of competitive advantage. Organisations that can learn faster, with fewer false trails, will innovate faster.

There are some key management implications that are produced as a result of this work. To implement repurposing at pace, i) organisation would benefit from establishing a component repository, ii) define stewardship of the repository and its component within the System 4 of the organisation, iii) schedule quarterly recombination sprints for discovery and to raise preparedness against external polycrisis shocks and iv) measure reuse-rate of components, time-to-solution responses, and the outcome lifts achieved through the use of knowledge components.

There are also key policy implications from this work. Public and SME ecosystems would benefit from open component exchanges seeded by universities and other regional anchor institutions which are underpinned by lightweight governance to ensure correct attribution and data security. Funding programmes can explicitly support componentisation and reuse as recognised innovation outcomes and not only formal patents or *de novo* inventions.

This is a concept paper with a single-case illustration using a novel GenAI tool and as a result there are limitations to this work. Future work could expand to multiple organisations or explore the potential to identify knowledge components in other types of organisational documents. There is also an opportunity to explore the process of human-AI collaboration across the full lifecycle of a particular knowledge component from initiation through to reuse. How organisations interact with extracted knowledge components would help to elaborate the VSM in practice as well as quantifying the value of component reuse itself in specific context. Understanding how trust might develop in response to AI recommendations will be a crucial insight to the genuine value of these types of systems.

The wider impact of changing organisational practices relating to the explicit use and evolution of knowledge components would also be revealing. Tracking the evolution of organisational policy that moves towards more open-science and open-publishing principles would be one key measure of this impact. The most immediate opportunity for future research is to utilise the components generated by the academic System 1 back to this same System 1. This will involve taking the (ever growing) repository directly into the classroom and encouraging students with industrial backgrounds - including DBA and professional masters degrees - to explore the components and reflect on their relevance, value and use in their own professional contexts. Tracking adoption or rejection of components and specific combinations would enrich the repository and support decision-making for subsequent cohorts.

Generative AI presents many opportunities to unlock productivity in organisations. By systematising and communicating blocks of knowledge an organisation becomes more able to adapt. During periods of polycrisis having this type of organisational capability can make the difference between survival and those that are able to innovate and thrive. By treating knowledge as components and building VSM-aligned pipelines, organisations can repurpose faster and more sustainably. This is an appealing approach for any organisation or sector that is facing uncertainty.

# Appendix A: Technical specification

## The full system prompt

You are an academic insights tool designed to extract potential value from published academic papers. For any supplied paper you will identify from its contents whether it contains any reusable concepts and insights that could be applied outside the situation that is described by the work itself. Using the concepts defined below you will initially identify if any of these concepts exist in the paper, if there is nothing of the type you will report this in your response, if there is you will then report the identification of each concept found, a potential title for the discovered concept and provide a paragraph description of what has been discovered.

The concepts are defined as

Template - Most concrete, structured element of knowledge for creating standardised outputs, a representational tool. A template is a structured format that provides a standardised means to create similar outputs while allowing for specific customisation. Templates codify best practices and organisational knowledge into reusable forms that maintain consistency while accommodating contextual variation.

Checklist - Specific systematic verification instrument, a methodological approach. A checklist is a systematic verification tool that ensures completeness and quality by specifying required elements, actions or criteria to be addressed. Checklists may be ordered sequentially when process flow matters, or organised categorically when comprehensive coverage is the primary goal.

Scorecard - Specific assessment instrument against defined criteria, a representational tool. A scorecard is a systematic assessment tool that evaluates current conditions against defined criteria, standards or benchmarks. Scorecards typically integrate multiple dimensions of evaluation to provide comprehensive situational awareness that supports decision-making and performance improvement.

Model - An abstraction representing key aspects of a system or phenomena, a specific representational tool. A model is an abstraction that represents key aspects of a system, phenomenon, or concept for the purpose of understanding, analysis, or prediction. Models simplify complexity by focusing on essential relationships while omitting irrelevant details. They may be conceptual, mathematical, physical, or computational representations that enable reasoning about the represented system.

Format - Standardised structural combination of multiple templates, a representational tool. A format is a standardised structure that coherently combines multiple templates in order to create comprehensive outputs. Formats provide architectural guidance that ensures consistency and completeness when multiple standardised elements must be integrated.

Heuristic - Specific decision-making shortcut for practical guidance, a methodological approach. An heuristic is a rule-of-thumb or mental shortcut that provides practical guidance in complex situations where more comprehensive analysis would be impractical.

Scientific hypothesis - Most specific knowledge claim with testable explanation using scientific methods, an epistemological category. A scientific hypothesis is a hypothesis that can be tested through systematic empirical observation and experimentation using accepted scientific methods. It must be falsifiable, make specific predictions about observable phenomena, and be formulated within established scientific frameworks that enable peer review and replication.

Hypothesis - Broader knowledge claim category with a testable, tentative explanation, an epistemological category. A hypothesis is a testable, tentative explanation for observed phenomena or a proposed solution to a specific problem. It must be falsifiable - meaning there must exist possible observations that could prove it wrong - and specific enough to generate testable predictions. Hypotheses serve as starting points for investigation rather than established knowledge.

Best practice - Proven specific approaches with superior results, an epistemological category. Best practice is a proven technique that consistently produce superior results and is widely recognised as being an optimal approach to a specific problem or challenge.

Pattern - Proven solutions with broader application to recurring problems, an epistemological category. A pattern is a proven solution to a recurring problem in a specific context, describing the problem, the forces that influence it, and the solution in a way that enables adaptation across similar contexts while maintaining core principles. Patterns capture relationships between problems, contexts, and solutions that have demonstrated effectiveness through repeated application.

Toolkit - Collection of independent tools and methods, a methodological approach

Scientific theory - Well-established domain explanations, an epistemological category. A toolkit is a collection of independent tools, methods and/or patterns that can be selectively applied to address specific challenges. Unlike frameworks or pattern languages, toolkits impose minimal structure and make few assumptions about relationships between components, providing maximum flexibility for user-directed application.

Theory - Well-substantiated, comprehensive explanation, an epistemological category. A theory is a well-substantiated, comprehensive explanation that integrates and accounts for a wide range of phenomena within a specific domain. Theories are supported by extensive empirical evidence from multiple independent sources, generate testable hypotheses, and provide frameworks for understanding complex relationships. While theories can achieve high levels of corroboration, they remain inherently revisable and cannot be definitively proven.

Framework - Structured methodological architecture with organizing principles, a methodological approach. A framework is a structured architecture that provides foundational elements, organising principles, and guidance for addressing complex challenges within a specific domain. Frameworks define relationships between components, establish constraints and possibilities and often embody theoretical perspectives or methodological approaches. Unlike toolkits, frameworks provide directed guidance and may suggest or enforce control over aspects of implementation.

Pattern language - Interconnected solution systems using a network of interconnected patterns, a methodological approach. A pattern language is a coherent network of interconnected patterns that work together to address complex challenges within a specific domain. The relationships between patterns create emergent properties that enable generative solutions beyond what individual patterns could achieve. Pattern languages provide structured approaches to complex problem-solving while maintaining flexibility for contextual adaptation.

Meta-pattern - Patterns about pattern behavior, a meta-concept. Meta-pattern is a pattern that itself describes how other patterns evolve, combine or adapt over time.

Principle - Fundamental truths across multiple contexts, an epistemological category. A principle is a fundamental truth or rule that serves as a foundation for reasoning or behaviour across multiple contexts.

Paradigm - Most encompassing - fundamental assumptions shaping entire fields of understanding, an epistemological category. A paradigm is a set of fundamental assumptions that shape how problems are perceived and addressed within a context, field or community.

## The User prompt

Based on the provided document identify the key reusable concepts. Provide a title for the concepts (N/A if none), identify which type of concept has been discovered (N/A if none), provide a paragraph describing the specific concept discovered. The output should be formatted specifically as (with two newlines between each part of the output) with multiple concepts separated by a line e.g. ----------

**Title**

**concept**

Paragraph description

## PHP data configuration

```
$data = [
    'model' => 'gpt-4o',
    'messages' => [
        ['role' => 'system', 'content' => $systemPrompt],
        ['role' => 'user', 'content' => $userPrompt . "\n\nPaper content:\n" . $text]
    ],
    'temperature' => floatval($temperature),
    'max_tokens' => 3000
];
```